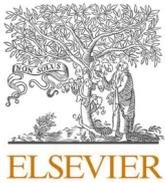
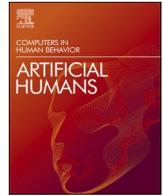

# Can ChatGPT read who you are?

Erik Derner [a,c,*], Dalibor Kučera [b], Nuria Oliver [a], Jan Zahálka [c]

[a] *ELLIS Alicante, Distrito Digital 5 (Puerto de Alicante) – Edificio A, Muelle de Poniente 5, 03001, Alicante, Spain*
[b] *Department of Psychology, Faculty of Education, University of South Bohemia, Jeronýmova 10, 37115, České Budějovice, Czech Republic*
[c] *Czech Institute of Informatics, Robotics, and Cybernetics, CTU in Prague, Jugoslávských partyzánů 1580, 16000, Prague, Czech Republic*

A B S T R A C T

The interplay between artificial intelligence (AI) and psychology, particularly in personality assessment, represents an important emerging area of research. Accurate personality trait estimation is crucial not only for enhancing personalization in human-computer interaction but also for a wide variety of applications ranging from mental health to education. This paper analyzes the capability of a generic chatbot, ChatGPT, to effectively infer personality traits from short texts. We report the results of a comprehensive user study featuring texts written in Czech by a representative population sample of 155 participants. Their self-assessments based on the Big Five Inventory (BFI) questionnaire serve as the ground truth. We compare the personality trait estimations made by ChatGPT against those by human raters and report ChatGPT's competitive performance in inferring personality traits from text. We also uncover a 'positivity bias' in ChatGPT's assessments across all personality dimensions and explore the impact of prompt composition on accuracy. This work contributes to the understanding of AI capabilities in psychological assessment, highlighting both the potential and limitations of using large language models for personality inference. Our research underscores the importance of responsible AI development, considering ethical implications such as privacy, consent, autonomy, and bias in AI applications.

## 1. Introduction

Advancements in artificial intelligence (AI), particularly in the analysis and generation of natural language, have revolutionized human-computer interaction. This AI-based revolution has been led by large language models (LLMs), such as ChatGPT, developed by OpenAI. ChatGPT stands out due to its unprecedented adoption rate and its exceptional ability to engage in coherent, contextually relevant conversations across diverse knowledge domains. Beyond its conversational capabilities, ChatGPT has shown potential in tasks including creative writing, question-answering, programming, language translation, text summarization, and problem-solving (Feng et al., 2023; Kocoń et al., 2023; Meyer et al., 2023; Sahari et al., 2023).

As human interaction with chatbots becomes more frequent and complex, the potential of LLMs to serve as tools for user modeling and communication analysis increases. Extensive research has demonstrated that personality traits can be reliably inferred from linguistic styles, suggesting significant implications for personalized interactions in human-computer interfaces (Ireland & Mehl, 2014; Pennebaker & King, 1999). While this demonstrates a promising application of LLMs, their use in personality assessment through language analysis has been relatively under-explored, especially outside of English-speaking contexts.

In fact, despite the advancements in AI and its integration into daily life, there remains a critical gap in understanding and applying these technologies for personality analysis across diverse linguistic and cultural backgrounds. This paper seeks to bridge this gap by exploring the capabilities of ChatGPT to automatically infer personality traits from text in a language other than English, specifically Czech – a West Slavic language spoken by more than 13 million people (Simons & Fennig, 2017). This focus addresses the scarcity of research regarding the use of LLMs in languages other than English (particularly in mid-to low-resource languages) and contributes to the broader goal of promoting inclusivity and fairness in AI applications.

By analyzing data collected by means of a user study, we assess the accuracy of ChatGPT in personality trait estimation using texts written by a representative sample of the Czech-speaking population and comparing these results to those obtained from human raters. In addition to measuring the capabilities of ChatGPT in a new domain, we reflect on both the potential and the challenges related to leveraging AI tools for psychological assessment.

Our study is positioned at the intersection of natural language processing (NLP), psychology, and linguistics, aiming to expand the current understanding of LLMs' abilities to analyze personality through language. It seeks to demonstrate how AI can be leveraged to enhance the personalization of human-computer interaction and to foster a more inclusive approach by considering diverse languages and cultural contexts.

## 2. Related work

The intersection between automatic natural language processing methods and psychology is an emerging field of study, focusing on understanding and interpreting different aspects of human traits and behavior through language technologies (Boyd & Schwartz, 2021). Exemplary applications include using AI algorithms to assess sentiment from text (social media posts or personal blogs) and identify potential indicators of mental health issues (Le Glaz et al., 2021; Neethu & Rajasree, 2013), to identify language patterns that may predict risky behaviors (Singh et al., 2020), to analyze the dynamics of social interactions (Hoey et al., 2018; Lane, 2013), to support in the diagnosis of mental health disorders (Corcoran & Cecchi, 2020) or to automatically assess personality traits (Jayaratne & Jayatilleke, 2020).

Regarding personality, prior research in psychology has established the link between individual linguistic patterns and personality traits: language has been found to provide significant insights into an individual's personality (Ireland & Mehl, 2014; Pennebaker & King, 1999), suggesting the potential of leveraging natural language processing (NLP) tools to automatically infer personality. Recent research has analyzed short communication intentions (Ramírez-de-la Rosa et al., 2023), consumed textual content (Sutton et al., 2023), or typing dynamics (Buker & Vinciarelli, 2023; Kovačević et al., 2023) to infer personality traits. However, applying LLMs to automatically assess personality traits is still in its incipient stages. Early work in this domain has primarily focused on straightforward text analysis tasks, such as sentiment analysis (Bu et al., 2023). Preliminary studies indicate that LLM-based chatbots, such as ChatGPT, exhibit promising potential in inferring personality traits from English text (Amin et al., 2023).

In this paper, we contribute to the interdisciplinary research on personality analysis using language-based AI models in four ways. First, we explore how well a generic chatbot (ChatGPT) built on top of a large language model (GPT-3.5) can infer its users' personality from short textual input. The ability to tailor responses based on inferred personality traits can enhance user experience by enabling personalization. Second, we contribute to the growing body of literature at the intersection of artificial intelligence and psychology. Such a multidisciplinary approach holds promise for refining AI models to better align with human cognitive processes and for building AI tools to support psychology researchers and practitioners. Third, we extend the body of research on languages with fewer resources by performing the study in Czech, a West Slavic language with approximately 13.2 million speakers (Simons & Fennig, 2017). Since the majority of studies involving large language models do not address the linguistic diversity, with this research, we contribute to the broader goal of promoting inclusivity and fairness in AI applications. Finally, we draw considerations on the potentialities and pitfalls of deploying large language models in real-world scenarios. While automatically inferring personality from text enables personalized and engaging user experiences, it also raises ethical concerns related to human autonomy consent, privacy, and biases. Through this study, we aim to shed light on these considerations and stimulate discussions on responsible AI development.

## 3. Method

In this section, we detail the dataset used, the experimental setup, and the evaluation metrics that underpin our analysis.

### 3.1. Dataset

We analyze the data collected by the psychological-linguistic project CPACT (Kučera, 2020; Kučera et al., 2022) focused on identifying personality markers in human-written text. Quota sampling ensured that the sample was comparable with the characteristics of the population in the Czech Republic in the categories of gender, age, and education level. The data of $N = 155$ individuals over 15 years of age (77 men, 78 women) were analyzed in this study.

The participants were administered the self-report Big Five Inventory (BFI-44) (John et al., 2008) to assess their characteristics. The BFI-44 is a 44-item questionnaire measuring five personality dimensions: extraversion (8 items), agreeableness (9 items), conscientiousness (9 items), neuroticism (i.e., emotional lability, 8 items), and openness to experience (10 items). It has favorable psychometric properties, represented by adequate scale reliability and corresponding retest reliability (Li et al., 2015). The Czech BFI-44 version was analyzed in adolescent and adult populations with a reliability spanning between 0.65 and 0.83. The approximate test–retest stability of BFI-44 dimensions after two months is $r = 0.79$ (Hřebíčková et al., 2016).

On the same day of collecting self-report questionnaires, the BFI-44 questionnaire was also filled out by another person, referred to as the *partner* in this paper, with whom the participant reported having a frequent and sincere relationship. The partner-assessment scores provides a valuable complement to the self-assessment and adds information on possible trends and asymmetries in the human assessment (Vazire, 2010).

Subsequently, the participants were asked to compose four short texts (letters) in their native Czech language with an overall length of 180–200 words each. All letters were typed on a computer on the same day, with mandatory breaks between writing texts. Participants were required to follow the described scenarios (L1–L4) summarized in Table 1, with an emphasis on the authenticity and realism of the communication.

Two human text raters were asked to assess the personality of all participants based on the provided texts. The raters were a female, aged 65 (rater A) and a male, aged 35 (rater B). Both hold a university degree (non-psychology), and they were trained to understand the construct of

**Table 1**
Instructions for participants to write four short letters (L1–L4). The instructions were given to the study participants in Czech; the table reports their English translation.

| Cover letter (L1) | Letter from vacation (L2) | Complaint letter (L3) | Letter of apology (L4) |
| --- | --- | --- | --- |
| You have found a job offer that captivated your interest, and you really aspire to be hired for this particular position. Therefore, you are going to write a letter to the company's director in response to his/her offer to try to persuade the director that you are the right candidate for this position. | You are enjoying your time on an amazing vacation. Everything is going well, as expected, and you are fully indulged in some popular activities. Therefore, you have decided to write a letter to your friend and convince him/her to come over and enjoy this perfect time with you. | Until recently, you were living contentedly in your apartment (house). Nevertheless, recently, issues arose that made your happy home more like a hellish home. Although you originally strived to sort out these issues in a gentle way, your efforts did not make any difference. Therefore, you decided to write an official complaint letter to the respective authorities. | You have done something that substantially harmed your relationship with someone you were close to for a long time. You promised something that you did not fulfill. You feel sorry, and you know that you made a mistake. Because you do not want to lose this person, you have decided to write a letter of apology to him/her. |

Big Five personality traits. They were instructed to read each text and then use a scale to estimate the degree of presence of each of the five personality characteristics, which they attribute to the author of the text. For the estimate, they used a three- or five-point scale ranging from 0 (characteristic not present) to 2 or 4 (dominant characteristic). The details on the range for each dimension are reported in Table 2. For example, for the agreeableness dimension, 0 corresponds to 'very aloof' and 4 corresponds to 'very warm-hearted' (Kučera et al., 2018).

As a result, the dataset consists of a total of $4 \times 155 = 620$ texts together with the assessment of the Big Five personality traits for each of the 155 participants provided by themselves (ground truth), their partner, and two human raters. In the following, we will refer to the scores from this study as follows:

- $A_S$ – personality *self-assessment* of the study participant using the standardized BFI-44 test (self-report variant);
- $A_P$ – participant's personality assessment by their *partner* using the standardized BFI-44 test (other-report variant);
- $H_A$, $H_B$ – personality estimation score based on the text evaluation by *human raters* A and B, respectively.

### 3.2. Automated personality estimation

In our experimental evaluation, we asked ChatGPT to score the letters exactly in the same way as the human raters were asked to score them. We created a set of Python scripts leveraging OpenAI's API access to the GPT-3.5 chat model (commonly known as ChatGPT), using the March 1, 2023 model version. The Hugging Face Transformers library was used to facilitate the interaction with ChatGPT.

To assess the BFI personality traits from the participants' letters without fine-tuning the model, we employed zero-shot prompting (Ziems et al., 2023). The principle of zero-shot prompting consists in providing all the needed context directly in the prompt. Zero-shot prompting allows us to leverage the power of ChatGPT's language comprehension abilities without modifying the model through additional training.

We experimented with four distinct prompt variants, each written in the Czech language, to investigate their impact on ChatGPT's performance in personality trait assessment. Each prompt consisted of at most three elements:

- **Task (T)** that specifies the output ChatGPT should provide, such as estimating the author's extraversion on a scale from 0 to 4, where 0 represents a strongly introverted person and 4 indicates a strongly extraverted person, requesting a single integer response;
- **Letter (L)**, which consists of the original letter written by the participant;
- **Dimension description (D)**, explaining the BFI trait according to its psychological definition (John et al., 2008).

Using these components, the prompts were constructed in four different ways:

- $GPT_{TL}$: Task + Letter;
- $GPT_{DTL}$: Dimension description + Task + Letter;
- $GPT_{LT}$: Letter + Task;
- $GPT_{DLT}$: Dimension description + Letter + Task.

These prompts aim to leverage ChatGPT's language comprehension capabilities to infer personality traits based on the provided letters (and the descriptions of the personality dimensions, where applicable). The four variants of the prompts were evaluated to understand the impact that different prompts have on the performance of the chatbot. As the prompts and the letters were in Czech, we empirically evaluated ChatGPT's capability to perform the task in a language with fewer resources.

Each of the 620 letters was treated by ChatGPT as a separate case, with no information about their authors or possible interconnection (e.g., such that each participant produced four texts). ChatGPT was asked to assess only one personality dimension at a time. To mitigate the influence of ChatGPT's stochastic nature, each prompt execution was performed five times. We report the mean of the five responses rounded to the nearest integer as a score assigned to the letter by ChatGPT. Occasionally (in less than 0.1 % cases), ChatGPT did not follow the instructions to return a single integer and delivered a wordier response instead. Such results were discarded, and the mean was calculated on the remaining valid values.

### 3.3. Evaluation metrics

The starting point in evaluating the success rate of the ChatGPT personality estimation was the premise that the scores obtained through the standardized self-assessment questionnaire ($A_S$) are the most accurate in characterizing the actual personality of the text authors, and hence we consider them as the ground truth. This premise is extensively supported by previous work (Funder, 1995; Mehl et al., 2006; Paulhus & Vazire, 2007; Pronin et al., 2001). This basis was also applied to the partner-assessments ($A_P$), which we present in this study as supporting information about the validity of the test method.

To allow for a comparison of all evaluation scores, we re-scaled the data as follows. The human raters **H** assigned an integer score to each letter using the range reported in Table 2. ChatGPT was instructed the same way as the human raters, i.e., it was given the same scale for each personality trait. The scale used by the human raters (and ChatGPT) served as the reference scale and all other personality assessment scales were transformed to match that scale. To that end, the original scores of the BFI-44 self-assessment $A_S$ and partner-assessment $A_P$ were transformed into equally-sized bins, corresponding to the aforementioned integer scale for each personality trait.

To determine the similarity between two or more assessments, a combination of methods and procedures is commonly used (Carlson & Kenny, 2012; Ledermann & Kenny, 2017). For the purpose of this study, we chose several methods: the root mean square error (RMSE), the mean absolute error (MAE), the hit rate, the F1 score, and Spearman's correlation coefficient ($\rho$). These methods are complemented by descriptive statistics and visual transformation of values (score spectra).

The RMSE is the square root of the mean squared differences between one type of assessment (human rater **H** or ChatGPT **GPT**) and the self-assessment $A_S$. The MAE is the average of the absolute differences between one type of assessment (human rater **H** or ChatGPT **GPT**) and the self-assessment $A_S$.

To compute the hit rate, we first labeled the personality score in each dimension as low, neutral, or high, as per Table 2. The hit rate measures the agreement between one type of assessment (human rater **H** or ChatGPT **GPT**) and the self-assessment $A_S$. When comparing two low/neutral/high scores, this metric aims to simplify the match result into a binary form of divergence/congruence of both assessments. For example, a participant who scored 3 in the BFI-44 extraversion dimension would be considered to be *high* in the extraversion dimension. If ChatGPT reported a value of 3 or 4, there is a hit because both assessments score the person as *high* in extraversion. We report both the absolute number of matches or hits and the hit rate as a percentage for low and high spectra. The hit rate is an easy-to-understand representation of the agreement between two methods (Fletcher, 2013; Haslam et al., 2020).

**Table 2**
Ranges and score values for each of the five personality dimensions. To unify the scales, the scores are grouped into low, neutral, and high spectra.

| Dimension | Range | Score spectra | | |
|---|---|---|---|---|
| | | Low | Neutral | High |
| Extraversion | 0–4 | 0, 1 | 2 | 3, 4 |
| Agreeableness | 0–4 | 0, 1 | 2 | 3, 4 |
| Conscientiousness | 0–2 | 0 | 1 | 2 |
| Neuroticism | 0–4 | 0, 1 | 2 | 3, 4 |
| Openness | 0–2 | 0 | 1 | 2 |

The F1 score, a widely utilized metric in binary classification, was adopted to measure the precision and recall balance between the assessments of the human rater **H** or ChatGPT **GPT** against the self-assessment $\mathbf{A_S}$. Precision is the ratio of correctly identified positive cases to all cases identified as positive and recall is the ratio of correctly identified positive cases to all actual positive cases. Specifically for our application, the high spectrum of each dimension, as defined in Table 2, represents the positive class (e.g., an extraverted person), while the neutral and low spectra correspond to the negative class (e.g., a person not characterized as extraverted). The F1 score captures the harmonic mean of precision and recall, providing a single measure of a method's accuracy in identifying high scores in each of the personality dimensions.

Finally, the Spearman's correlation coefficient ($\rho$) was used to determine the degree of association between the assessments of personality traits given by the human rater **H** or ChatGPT **GPT** on one side and the self-assessment $\mathbf{A_S}$ on the other side. This non-parametric measure is particularly suitable for our analysis as it measures how well the relationship between two assessments can be described using a monotonic function, thus providing insight into the consistency of orderings between different types of assessments. It is appropriate for comparing ordinal data, like the personality scores in our study.

## 4. Results

In this section, we report the general descriptives and the RMSE and MAE metrics, the score spectra, the hit rate, and the F1 score. This descriptive part is followed by the results of inferential statistics – Spearman's correlation coefficients.

### 4.1. General descriptives

General descriptives of the self-assessment and all the evaluation data are presented in Table 3. Key to the interpretation of the table is the first column $\mathbf{A_S}$, which contains the participants' self-assessment that we consider as ground truth. Scores obtained by other methods should be as close as possible to these values.

An interesting observation can be made with respect to the coefficient of variation. It shows how the values vary across all assessments (evaluations). For example, in the extraversion dimension, the variability of the ChatGPT evaluation (**GPT**) is considerably lower than that of the $\mathbf{A_P}$/$\mathbf{H_A}$/$\mathbf{H_B}$ assessments. This means that the ChatGPT assessments do not cover the full spectrum of values that characterize this dimension in the ground truth (self-assessment scores $\mathbf{A_S}$).

### 4.2. RMSE and MAE

Table 4 reports the RMSE and MAE metrics that measure the proximity of the assessment scores to the self-assessment mean. To interpret the metrics correctly, note that conscientiousness and openness use a smaller scale than the other dimensions (see Table 2).

As expected, the personality estimation of the participant's partner $\mathbf{A_P}$ was the most accurate, followed by that of ChatGPT. Interestingly, both human raters were less successful than ChatGPT in inferring the author's personality from the letter, especially $\mathbf{H_A}$. ChatGPT outperforms both human raters by the most significant margin in the $\mathbf{GPT_{TL}}$ variant. Only for the conscientiousness dimension, **GPT** was outperformed by $\mathbf{H_B}$ in terms of RMSE. Overall, the other three **GPT** variants also achieved better average results than human raters.

**GPT**'s estimations were the most accurate in the agreeableness dimension. In terms of RMSE and MAE, all **GPT** variants perform better than the human raters on this trait and even outperform the partner's assessment $\mathbf{A_P}$.

**Table 3**
Descriptive statistics of all assessment scores. Letter evaluations (**H** and **GPT**) consist of 620 values each. Self- and other-assessment scores (**A**) comprise 155 values each.

|  | $A_S$ | $A_P$ | $H_A$ | $H_B$ | $GPT_{TL}$ | $GPT_{DTL}$ | $GPT_{LT}$ | $GPT_{DLT}$ |
|---|---|---|---|---|---|---|---|---|
| **Extraversion** | | | | | | | | |
| Mode | 3 | 3 | 3 | 2 | 3 | 3 | 3 | 3 |
| Median | 2 | 3 | 3 | 2 | 3 | 3 | 3 | 3 |
| Mean | 2.194 | 2.523 | 2.350 | 2.153 | 2.694 | 2.903 | 2.652 | 2.865 |
| Std. deviation | 1.030 | 0.993 | 1.046 | 0.788 | 0.548 | 0.441 | 0.702 | 0.516 |
| Coeff. of variation | 0.470 | 0.394 | 0.445 | 0.366 | 0.203 | 0.152 | 0.265 | 0.180 |
| **Agreeableness** | | | | | | | | |
| Mode | 3 | 3 | 3 | 2 | 3 | 3 | 3 | 3 |
| Median | 3 | 3 | 3 | 2 | 3 | 3 | 3 | 3 |
| Mean | 2.729 | 2.774 | 2.397 | 2.087 | 2.777 | 2.752 | 2.742 | 2.831 |
| Std. deviation | 0.822 | 0.832 | 0.972 | 0.867 | 0.523 | 0.571 | 0.641 | 0.553 |
| Coeff. of variation | 0.301 | 0.300 | 0.405 | 0.415 | 0.188 | 0.207 | 0.234 | 0.195 |
| **Conscientiousness** | | | | | | | | |
| Mode | 1 | 2 | 2 | 1 | 1 | 2 | 2 | 2 |
| Median | 1 | 1 | 2 | 1 | 1 | 2 | 2 | 2 |
| Mean | 1.284 | 1.471 | 1.479 | 1.079 | 1.145 | 1.466 | 1.690 | 1.692 |
| Std. deviation | 0.610 | 0.549 | 0.575 | 0.497 | 0.491 | 0.566 | 0.552 | 0.551 |
| Coeff. of variation | 0.475 | 0.373 | 0.389 | 0.461 | 0.428 | 0.386 | 0.327 | 0.326 |
| **Neuroticism** | | | | | | | | |
| Mode | 2 | 3 | 3 | 2 | 2 | 1 | 2 | 1 |
| Median | 2 | 2 | 3 | 2 | 2 | 1 | 2 | 2 |
| Mean | 1.961 | 2.239 | 2.592 | 2.211 | 1.811 | 1.489 | 2.058 | 1.640 |
| Std. deviation | 0.990 | 0.979 | 0.951 | 0.893 | 0.613 | 0.595 | 0.829 | 0.681 |
| Coeff. of variation | 0.505 | 0.437 | 0.367 | 0.404 | 0.339 | 0.399 | 0.403 | 0.415 |
| **Openness** | | | | | | | | |
| Mode | 2 | 2 | 1 | 1 | 1 | 1 | 2 | 1 |
| Median | 2 | 1 | 1 | 1 | 1 | 1 | 2 | 1 |
| Mean | 1.497 | 1.458 | 1.226 | 1.031 | 1.031 | 0.840 | 1.558 | 1.282 |
| Std. deviation | 0.526 | 0.571 | 0.671 | 0.386 | 0.229 | 0.567 | 0.592 | 0.667 |
| Coeff. of variation | 0.351 | 0.392 | 0.547 | 0.375 | 0.222 | 0.675 | 0.380 | 0.520 |

**Table 4**
RMSE and MAE metrics indicating distance to the self-assessment ($A_S$) values (lower is better). The best results among methods based on text evaluation are shown in bold. The partner-assessments ($A_P$) are included for reference.

|                   | $A_P$  | $H_A$  | $H_B$  | $GPT_{TL}$ | $GPT_{DTL}$ | $GPT_{LT}$ | $GPT_{DLT}$ |
|-------------------|--------|--------|--------|------------|-------------|------------|-------------|
| **RMSE**          |        |        |        |            |             |            |             |
| Extraversion      | 0.873  | 1.416  | 1.301  | **1.260**  | 1.318       | 1.313      | 1.326       |
| Agreeableness     | 1.028  | 1.259  | 1.345  | **0.940**  | 0.966       | 0.995      | 0.947       |
| Conscientiousness | 0.698  | 0.799  | **0.754** | 0.782   | 0.808       | 0.867      | 0.852       |
| Neuroticism       | 1.109  | 1.505  | 1.353  | **1.175**  | 1.238       | 1.256      | 1.242       |
| Openness          | 0.594  | 0.883  | 0.797  | **0.728**  | 0.995       | 0.774      | 0.840       |
| **MAE**           |        |        |        |            |             |            |             |
| Extraversion      | 0.635  | 1.077  | 0.994  | **0.965**  | 1.000       | 1.008      | 1.008       |
| Agreeableness     | 0.737  | 0.928  | 1.026  | 0.675      | 0.698       | 0.712      | **0.670**   |
| Conscientiousness | 0.474  | 0.564  | 0.681  | **0.551**  | 0.563       | 0.633      | 0.623       |
| Neuroticism       | 0.808  | 1.181  | 1.040  | **0.901**  | 0.960       | 0.970      | 0.978       |
| Openness          | 0.340  | 0.631  | 0.577  | **0.524**  | 0.750       | 0.532      | 0.593       |

*4.3. Score spectra*

The relative frequencies of scores in all three spectra (low, neutral, and high) of a given personality dimension are shown in Fig. 1. This evaluation aims to map which spectrum is generally preferred by each assessment method. Evaluation scores following the distribution of self-assessment ($A_S$) can be considered as a substantial condition for achieving accurate evaluation.

The bar charts indicate that ChatGPT exhibits a 'positivity bias' in all dimensions: it tends to evaluate people as extraverted, agreeable, conscientious, emotionally stable (i.e., with low neuroticism scores), and open to experience. This tendency will be further discussed in Section 5. Another noteworthy observation is that ChatGPT tends to use the neutral score much less frequently when the task is given at the end of the prompt (in variants $GPT_{LT}$ and $GPT_{DLT}$) as compared to specifying the task before the letter (in variants $GPT_{TL}$ and $GPT_{DTL}$). In other words, ChatGPT appears to be more confident in assessing personality if the task is provided at the end of the prompt.

Complete descriptive statistics of values in three spectra are presented in Tables 7–11, see Appendix. Each of the personality dimensions in the three spectra (low, neutral, and high score) is described in terms of central tendency and variability. The table provides valuable information about how far a given assessment method is from the values of referential self-assessment scores, from which we can infer the tendency and degree of specific bias.

*4.4. Hit rate*

Next, we evaluate the accuracy of all personality assessment methods by means of the hit rate. This metric analyzes the agreement between the self-assessment $A_S$ and each evaluation method. Table 5 reports the absolute number of matches and the corresponding percentage for the low and high spectra of each personality trait. It shows considerable variability in the accuracy of the assessments for all methods (**H**, **GPT**).

When comparing the performance, it is important to consider both ends of the spectra and their size. Similarly as with the RMSE and MAE metrics, **GPT** yields the best results on the agreeableness dimension in the high spectrum, which is almost ten times larger than the low spectrum, with a hit rate ranging from **84 %** to **91 %** depending on the prompt variant. It outperforms the human raters **H** and even the partner's assessment $A_P$. In the openness dimension, the results vary substantially between the **GPT** variants, implying that this personality trait is difficult for ChatGPT to infer. Prompt variants with the task specified at the end ($GPT_{LT}$, $GPT_{DLT}$) show better performance. For the remaining dimensions, the results are mixed: **GPT** performs in some cases better than other evaluators and in other cases worse.

These results support again the presence of a positivity bias in ChatGPT. Nevertheless, it still reflects the specificity of the author/text, which is particularly noticeable in the dimensions of agreeableness and conscientiousness. It is also noteworthy that among the successful **GPT** variants, the ones that yield the best results are in most cases those that include the personality trait description in the prompt ($GPT_{DTL}$, $GPT_{DLT}$).

*4.5. F1 score*

The F1 scores for the various evaluation methods across different personality dimensions are presented in Fig. 2. This metric, a harmonic mean of precision and recall, offers insights into the balance between the accuracy and completeness of each evaluation method.

In the extraversion and agreeableness dimensions, all **GPT** variants show consistently good results, outperforming both human raters **H**. Furthermore, all **GPT** variants outperform even the partner-assessment $A_P$ in the agreeableness dimension. Regarding conscientiousness, **GPT** exhibits performance comparable or superior to the performance of human raters **H**, except for the $GPT_{TL}$ variant. ChatGPT's positivity bias is evident when inferring neuroticism which yields low F1 scores as the chatbot avoids providing high scores for this trait. Finally, **GPT**'s performance for the openness dimension significantly depends on the variant. Specifying the task at the end of the prompt ($GPT_{LT}$ and $GPT_{DLT}$ variants) helps ChatGPT substantially improve its performance in terms of the F1 score. The lower performance of the human evaluators and particularly of $H_B$ underscores the challenges in the human judgment of openness from short texts.

*4.6. Correlation coefficients*

The results of inferential statistics in the form of Spearman's correlation coefficient were performed both for all text types (L1–L4) and for each text type individually (L1, L2, L3, and L4). Table 6 shows the correlation coefficients $\rho$ and their significance levels. The results indicate that the covariance of self-assessment scores ($A_S$) with ChatGPT scores (**GPT**) varies across personality dimensions and texts. To summarize the results, we report only relations that can be considered at least weakly correlated, which corresponds to $\rho > 0.2$ (Cohen, 2013).

Significant positive correlations were found for letter L4 (apology letter) and extraversion, for letter L3 (complaint letter) and agreeableness, and for letter L2 (letter from vacation) and conscientiousness. No significant correlations were found between the letter types and the ChatGPT scores for the neuroticism and openness dimensions.

If we were to compare ChatGPT's assessments with the human evaluations, ChatGPT achieved results comparable to human rater $H_A$ and outperformed human rater $H_B$. However, note that the correlation values are low and/or non-significant for both human raters and ChatGPT. Thus, the results rather indicate that neither form of evaluation was very successful in terms of their correlation with the self-assessments.

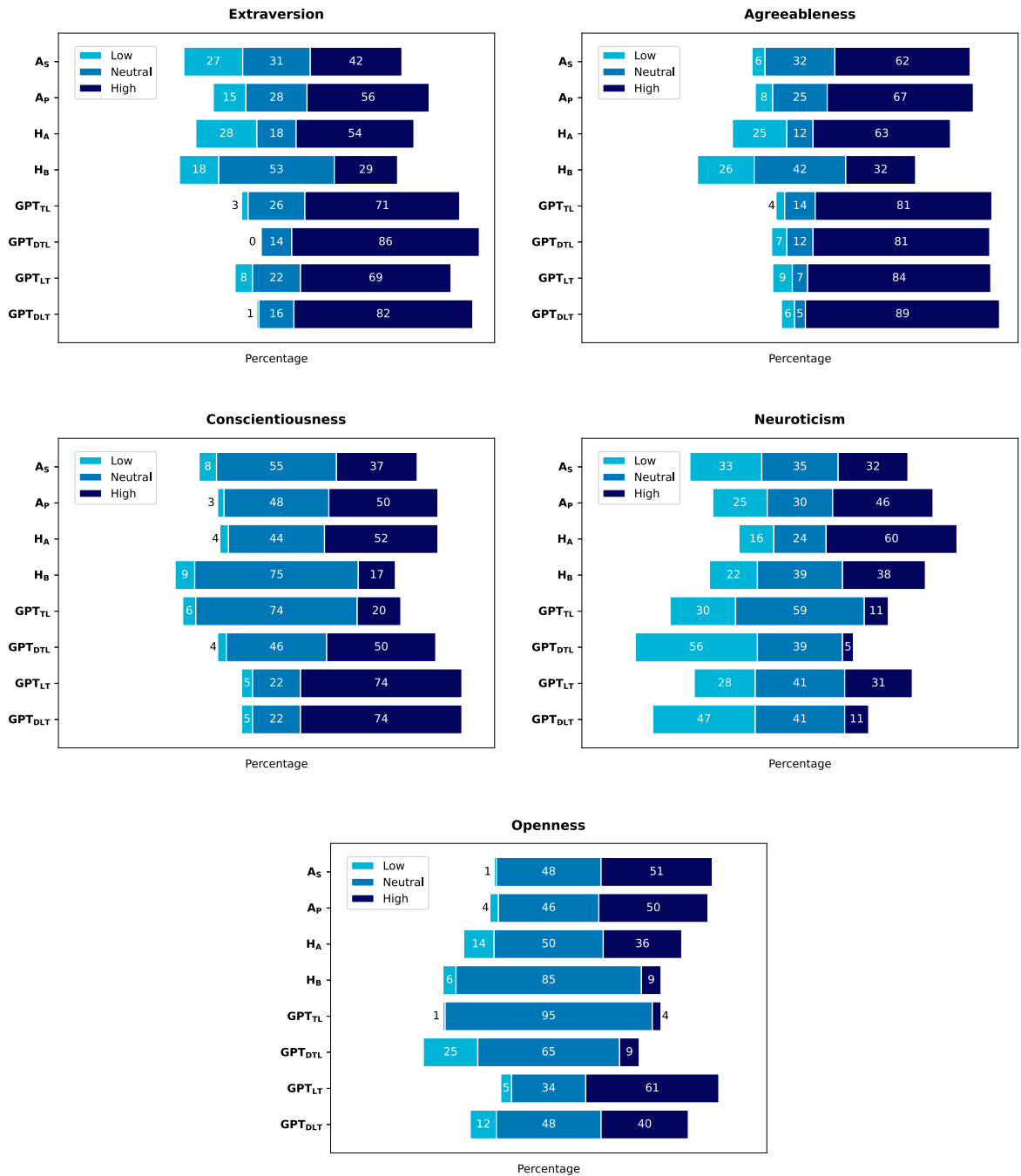

**Fig. 1.** Relative frequencies for all assessment scores in the low, neutral, and high spectra for all personality dimensions.

*4.7. Limitations*

Our findings demonstrate that ChatGPT, while showing competitive performance when compared to human raters, presents limitations in terms of absolute accuracy across all personality dimensions, particularly in neuroticism and openness. However, these results are significant as they illustrate the potential of LLMs in settings where traditional assessments might not be feasible or where a rapid, preliminary personality assessment is beneficial. Additionally, our analysis reveals that ChatGPT's performance is remarkably close to human raters in terms of RMSE, MAE, hit rate, F1 score, and correlation for several personality dimensions. This suggests a promising baseline capability of ChatGPT to operate alongside traditional measures.

## 5. Discussion

In this paper, we have empirically evaluated the capabilities of a general-purpose LLM-based chatbot, ChatGPT, to infer personality from short texts, and we have compared its performance with that of two human raters. Surprisingly, ChatGPT's assessments outperformed human assessments according to most metrics (RMSE, MAE, hit rate, F1 score, and correlation) in several personality dimensions, yet we also uncovered interesting findings that reveal the strengths and limitations of chatbots in inferring personality from text.

*5.1. Positivity bias*

We have identified a *positivity bias* in ChatGPT's assessments, i.e., its tendency to assign socially desirable scores across key personality dimensions. Social desirability is defined in psychology as the bias or tendency of individuals to present themselves in a manner that will be viewed favorably by others (American Psychological Association, 2023; Svoboda et al., 2001). Most authors agree that there are at least two levels of social desirability: (1) the level of self-deception (a reporter has a distorted self-image) and (2) the level of other-deception, i.e., deliberate deception of others or so-called impression management (Figurová, 2007). Both levels are related to typical motivational patterns (McFarland & Ryan, 2000). If we dare to speculate and project these psychological constructs into the ChatGPT processes, we could attribute the bias to (1) its pro-social naivety, i.e., ChatGPT is unintentionally mistaken, or to (2) a strategic pandering to the user. From a technical perspective, this positivity bias aligns with the inherent design of language models to favor more positive or neutral rather than negative content in their responses, particularly when they are fine-tuned by means of human feedback (Ouyang et al., 2022). Furthermore, a similar trend is found in human-to-human interaction with the so-called friendship bias (Wood et al., 2010), i.e., a situation where humans underestimate the undesirable characteristics of others as a manifestation or confirmation of the positivity of their relationship (Kučera, 2020).

**Table 5**
Agreement between the self-assessment score (A$_S$) and other evaluation methods in the low and high spectra for each dimension (higher is better). The best results among methods based on text evaluation are shown in bold. The partner-assessment (A$_P$) is included for reference.

|  | A$_P$ | H$_A$ | H$_B$ | GPT$_{TL}$ | GPT$_{DTL}$ | GPT$_{LT}$ | GPT$_{DLT}$ |
|---|---|---|---|---|---|---|---|
| **Extraversion** | | | | | | | |
| **Low** (N = 168) | | | | | | | |
| Num. of matches | 64 | **57** | 32 | 9 | 0 | 13 | 2 |
| Hit rate (%) | 38 | **34** | 19 | 5 | 0 | 8 | 1 |
| **High** (N = 260) | | | | | | | |
| Num. of matches | 228 | 152 | 81 | 183 | **227** | 182 | 216 |
| Hit rate (%) | 88 | 58 | 31 | 70 | **87** | 70 | 83 |
| **Agreeableness** | | | | | | | |
| **Low** (N = 40) | | | | | | | |
| Num. of matches | 16 | 10 | **12** | 2 | 3 | 6 | 4 |
| Hit rate (%) | 40 | 25 | **30** | 5 | 8 | 15 | 10 |
| **High** (N = 384) | | | | | | | |
| Num. of matches | 268 | 254 | 128 | 322 | 322 | 330 | **351** |
| Hit rate (%) | 70 | 66 | 33 | 84 | 84 | 86 | **91** |
| **Conscientiousness** | | | | | | | |
| **Low** (N = 52) | | | | | | | |
| Num. of matches | 12 | 6 | **12** | 8 | 5 | 5 | 5 |
| Hit rate (%) | 23 | 12 | **23** | 15 | 10 | 10 | 10 |
| **High** (N = 228) | | | | | | | |
| Num. of matches | 144 | 133 | 49 | 42 | 128 | 180 | **184** |
| Hit rate (%) | 63 | 58 | 21 | 18 | 56 | 79 | **81** |
| **Neuroticism** | | | | | | | |
| **Low** (N = 204) | | | | | | | |
| Num. of matches | 84 | 35 | 43 | 64 | **120** | 65 | 94 |
| Hit rate (%) | 41 | 17 | 21 | 31 | **59** | 32 | 46 |
| **High** (N = 200) | | | | | | | |
| Num. of matches | 140 | **120** | 81 | 19 | 11 | 62 | 21 |
| Hit rate (%) | 70 | **60** | 41 | 10 | 6 | 31 | 11 |
| **Openness** | | | | | | | |
| **Low** (N = 8) | | | | | | | |
| Num. of matches | 4 | 1 | 0 | 2 | **3** | 1 | 1 |
| Hit rate (%) | 50 | 13 | 0 | 25 | **38** | 13 | 13 |
| **High** (N = 316) | | | | | | | |
| Num. of matches | 220 | 120 | 30 | 15 | 36 | **201** | 140 |
| Hit rate (%) | 70 | 38 | 9 | 5 | 11 | **64** | 44 |

### 5.2. Prompt and text dependency

ChatGPT's performance is sensitive to the formulation of the *prompt*. Including the descriptions of the personality dimensions in the prompt (**GPT$_{DTL}$** and **GPT$_{DLT}$** variants) enhanced ChatGPT's accuracy, which suggests that providing an explicit context within the prompt can guide the LLM towards more accurate evaluations. Specifying the task at the end of the prompt (**GPT$_{LT}$** and **GPT$_{DLT}$** variants) improves ChatGPT's performance in terms of the score spectra, hit rate, and F1 score. This can be attributed to the attention mechanism used in transformer-based LLMs (Vaswani et al., 2017), which could prioritize information toward the end of the prompt.

In addition, we identified a dependency on the type of letter for the assessment of different personality traits: statistically significant correlations between the self-assessments and ChatGPT's assessments were found for apology letters in the case of extraversion, for complaint letters in the case of agreeableness and for vacation letters in the case of conscientiousness. There were no significant correlations in the neuroticism and openness dimensions, which suggests that these two personality traits are more difficult to infer from text. Previous work has also reported that there is an interaction between personality traits and different types of information in zero-acquaintance settings such that different personality traits may be accurately judged from different types of information (Küfner et al., 2010). Openness has been reported to be related to what people tell about themselves (Back et al., 2010). Accuracy in agreeableness judgments has been seldom found in the literature (Küfner et al., 2010), and scholars have claimed that it is not possible to accurately infer neuroticism from text when no self-related content is provided (Küfner et al., 2010), which is aligned with our findings. However, while previous work has argued that extraversion and conscientiousness are not revealed in the style of linguistic expression (Küfner et al., 2010), we find that they manifest themselves in some types of text, such as apology (extraversion) and vacation (conscientiousness) letters.

### 5.3. Variability in performance

The variability in ChatGPT's performance to infer different personality traits illustrates the complexity of inferring nuanced human characteristics from text alone. ChatGPT's success in assessing agreeableness and extraversion, for instance, contrasts with its difficulties in accurately evaluating neuroticism or openness, highlighting the challenge of capturing the full spectrum of human personality through automated linguistic analysis.

### 5.4. Ethical considerations

The interplay between machine learning and psychology, as demonstrated in this study, has significant implications for advancing our understanding of human behavior and cognition. The ability of ChatGPT to mirror human-like personality assessments from text opens new areas of research and applications in cognitive science. This integration offers insights into how language reflects underlying personality traits and psychological conditions. Moreover, it provides a framework for developing more personalized and adaptive AI systems that can better understand and interact with users based on their unique characteristics.

However, this capability also introduces the potential for manipulation. When AI systems understand and predict the personalities of their users, there exists a risk of exploiting these insights for manipulative purposes, such as targeted advertising, political campaigning, or social engineering attacks. Furthermore, such a capability raises additional concerns about user privacy and the potential for misuse of personal data. Thus, it is imperative to establish ethical guidelines and robust safeguards, such as explicit user consent for personality analysis, strict privacy controls, transparency, and clear boundaries on how personality insights might be used. AI systems that interact with humans, such as chatbots, should always be designed from a human-centric perspective, with a focus on ethical personalization, prioritizing user well-being and autonomy over commercial or political gains. The collaboration between AI researchers and cognitive scientists is crucial in this context. It can lead to the development of AI systems that are not only technically advanced but, more importantly, aligned with and respectful of human

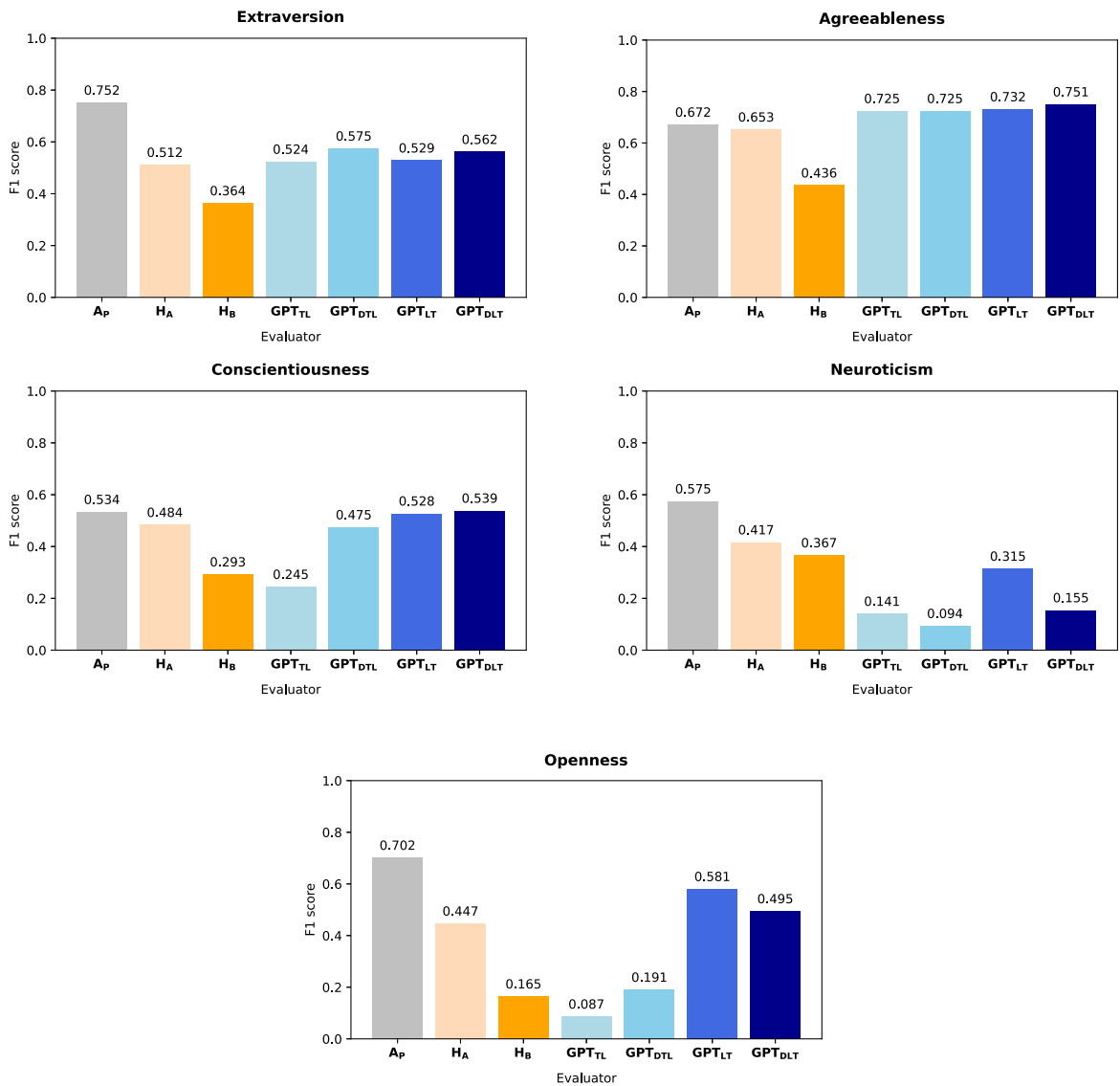

**Fig. 2.** F1 scores for all evaluation methods and all personality dimensions (higher is better). The self-assessment score (**A_S**) serves as the ground truth.

values, unlocking the potential of AI to support and augment—not replace—humans.

### 6. Conclusion

This study explores ChatGPT's ability to assess personality traits from short texts in the Czech language. It offers valuable insights into the performance of general-purpose LLM-based chatbots for psychological profiling. ChatGPT demonstrates a promising capability to automatically infer several personality dimensions from certain types of text, particularly extraversion, agreeableness, and conscientiousness. However, we have also identified limitations in ChatGPT's performance, such as a positivity bias, a dependency on the formulation of the prompt, and varying accuracy levels across different personality traits and text types. Our results highlight the potential of AI in supporting psychological assessments and interventions, emphasizing the importance of incorporating multi-disciplinary perspectives from AI and cognitive science.

While the current findings do not establish a superiority of ChatGPT over traditional assessment methods, they do demonstrate that ChatGPT's personality estimations align well with – or even outperform – those provided by human raters in terms of reliability and validity for certain personality traits. These findings underscore the predictive validity of ChatGPT in capturing personality dimensions that are discernible from linguistic patterns. Moreover, the application of ChatGPT in this context provides rapid and scalable assessments that provide an alternative to traditional methods. This is particularly useful in environments where quick screening is beneficial or where access to professional psychological assessment is limited.

From an ethical perspective, we underscore the need for cautious and responsible use of AI in personal and psychological assessments. The ethical implications related to privacy, consent, autonomy, and the potential for biases in automated personality analysis require further exploration and regulation. Ensuring transparency, safeguarding user data, preserving and honoring human autonomy, and mitigating biases are critical considerations as we integrate AI more deeply into personal and psychological domains. At the same time, ChatGPT's capabilities in inferring personality from text open the path for personalized interactions and enhanced user experience. Such personalization could help chatbots adapt to the preferred communication style of their users and increase user trust. These aspects could play a key role in the application of AI systems in psychological counseling and delivering psychological care.

In sum, while ChatGPT represents a significant step forward in AI's ability to analyze and interpret human language, its application in psychology-related domains requires careful consideration and further research. The interconnection of machine learning and cognitive science presents a promising direction that needs to be explored with caution and a commitment to ethical principles.

**Table 6**
Spearman's correlations $\rho$ between the self-assessment ($A_S$) and other evaluation methods (higher is better). Stars indicate levels of significance: one star denotes $p < 0.05$, two stars denote $p < 0.01$, and three stars denote $p < 0.001$. The partner-assessment ($A_P$) is included for reference. L1–L4 = 620 texts, L1/2/3/4 = 155 texts each.

| | $A_P$ | $H_A$ | $H_B$ | $GPT_{TL}$ | $GPT_{DTL}$ | $GPT_{LT}$ | $GPT_{DLT}$ |
|---|---|---|---|---|---|---|---|
| **Extraversion** | | | | | | | |
| L1–L4 | 0.697*** | 0.087* | 0.007 | 0.021 | 0.041 | 0.02 | 0.01 |
| L1 (Cover letter) | | 0.161* | −0.052 | 0.157 | 0.022 | 0.005 | 0.022 |
| L2 (Letter from vacation) | | 0.11 | 0.124 | 0.005 | −0.038 | −0.009 | −0.036 |
| L3 (Complaint letter) | | −0.026 | −0.075 | −0.164* | −0.02 | 0.082 | −0.056 |
| L4 (Letter of apology) | | 0.111 | 0.001 | 0.176* | 0.266*** | 0.084 | 0.168* |
| **Agreeableness** | | | | | | | |
| L1–L4 | 0.195*** | 0.101* | 0.02 | 0.09* | 0.093* | 0.09* | 0.099* |
| L1 (Cover letter) | | 0.141 | 0.1 | 0.002 | −0.023 | −0.053 | −0.043 |
| L2 (Letter from vacation) | | 0.2* | −0.004 | 0.012 | 0.087 | 0.013 | 0.047 |
| L3 (Complaint letter) | | 0.04 | −0.044 | 0.207** | 0.217** | 0.274*** | 0.279*** |
| L4 (Letter of apology) | | 0.151 | 0.014 | 0.093 | 0.08 | 0.049 | −0.016 |
| **Conscientiousness** | | | | | | | |
| L1–L4 | 0.293*** | 0.13*** | 0.148*** | 0.019 | 0.104** | 0.113** | 0.158*** |
| L1 (Cover letter) | | 0.135 | 0.119 | −0.058 | 0.099 | 0.083 | 0.051 |
| L2 (Letter from vacation) | | 0.248** | 0.093 | 0.33*** | 0.266*** | 0.283*** | 0.25** |
| L3 (Complaint letter) | | 0.108 | 0.126 | −0.051 | 0.133 | 0.018 | 0.068 |
| L4 (Letter of apology) | | 0.104 | 0.274*** | −0.079 | 0.027 | 0.147 | 0.251** |
| **Neuroticism** | | | | | | | |
| L1–L4 | 0.421*** | 0.000 | 0.009 | −0.005 | 0.03 | 0.053 | −0.001 |
| L1 (Cover letter) | | −0.045 | −0.014 | −0.051 | −0.003 | 0.028 | 0.043 |
| L2 (Letter from vacation) | | −0.08 | −0.082 | −0.111 | −0.029 | 0.008 | −0.064 |
| L3 (Complaint letter) | | 0.061 | 0.056 | −0.041 | 0.091 | 0.071 | −0.037 |
| L4 (Letter of apology) | | 0.061 | 0.096 | 0.113 | 0.125 | 0.204* | 0.059 |
| **Openness** | | | | | | | |
| L1–L4 | 0.42*** | 0.03 | 0.028 | 0.061 | 0.081* | 0.055 | 0.091* |
| L1 (Cover letter) | | 0.086 | 0.069 | 0.09 | 0.041 | 0.039 | 0.062 |
| L2 (Letter from vacation) | | −0.018 | 0.093 | 0.073 | 0.119 | 0.102 | 0.142 |
| L3 (Complaint letter) | | −0.022 | 0.025 | 0.024 | 0.125 | 0.044 | 0.182* |
| L4 (Letter of apology) | | 0.086 | −0.092 | 0.078 | 0.124 | 0.168* | 0.108 |

### CRediT authorship contribution statement

**Erik Derner:** Writing – original draft, Visualization, Software, Methodology, Investigation, Data curation, Conceptualization. **Dalibor Kučera:** Writing – original draft, Validation, Resources, Methodology, Formal analysis, Conceptualization. **Nuria Oliver:** Writing – review & editing, Supervision, Investigation, Conceptualization. **Jan Zahálka:** Writing – review & editing, Visualization, Methodology.

### Declaration of competing interest

The authors declare the following financial interests/personal relationships which may be considered as potential competing interests: Erik Derner reports financial support was provided by Intel Corporation. If there are other authors, they declare that they have no known competing financial interests or personal relationships that could have appeared to influence the work reported in this paper.

### Acknowledgements

The work of E.D. and N.O. on this paper was supported by the Valencian Government (Convenio Singular signed with Generalitat Valenciana, Conselleria de Innovacion, Industria, Comercio y Turismo, Direccion General de Innovacion). D.K.'s work was supported by the Fulbright-Masaryk Scholarship (reg. no. 2020-28-11) and by the Czech Science Foundation (reg. no. 16-19087S). E.D.'s work was also supported by Intel Corporation. J.Z.'s work was co-funded by the European Union under the project ROBOPROX (reg. no. CZ.02.01.01/00/22_008/0004590).

### Appendix A. Descriptives of the Score Spectra

Descriptives of the evaluation scores in three spectra in relation to the self-assessment reference score $A_S$ are shown in Table 7 for extraversion, in Table 8 for agreeableness, in Table 9 for conscientiousness, in Table 10 for neuroticism, and in Table 11 for openness.

**Table 7**
Descriptives of the evaluation scores in three spectra (low = 1, neutral = 2, high = 3) in relation to the self-assessment reference score $A_S$ for the extraversion dimension. Note that whenever more than one mode exists, only the first one is reported.

| Extraversion | $A_S$ | $A_P$ | $H_A$ | $H_B$ | $GPT_{TL}$ | $GPT_{DTL}$ | $GPT_{LT}$ | $GPT_{DLT}$ |
|---|---|---|---|---|---|---|---|---|
| **Low** (N = 168) | | | | | | | | |
| Mode | 1 | 2 | 3 | 2 | 3 | 3 | 3 | 3 |
| Median | 1 | 2 | 2 | 2 | 3 | 3 | 3 | 3 |
| Mean | 1 | 1.69 | 2.137 | 2.137 | 2.619 | 2.798 | 2.595 | 2.792 |
| Std. Deviation | 0 | 0.599 | 0.895 | 0.709 | 0.587 | 0.403 | 0.631 | 0.436 |

(*continued on next page*)

**Table 7** (*continued*)

| Extraversion | $A_S$ | $A_P$ | $H_A$ | $H_B$ | $GPT_{TL}$ | $GPT_{DTL}$ | $GPT_{LT}$ | $GPT_{DLT}$ |
|---|---|---|---|---|---|---|---|---|
| Coeff. of var. | 0 | 0.354 | 0.419 | 0.332 | 0.224 | 0.144 | 0.243 | 0.156 |
| **Neutral** (N = 192) | | | | | | | | |
| Mode | 2 | 3 | 3 | 2 | 3 | 3 | 3 | 3 |
| Median | 2 | 3 | 3 | 2 | 3 | 3 | 3 | 3 |
| Mean | 2 | 2.438 | 2.271 | 2.036 | 2.74 | 2.88 | 2.594 | 2.813 |
| Std. Deviation | 0 | 0.706 | 0.85 | 0.666 | 0.474 | 0.341 | 0.68 | 0.442 |
| Coeff. of var. | 0 | 0.29 | 0.374 | 0.327 | 0.173 | 0.118 | 0.262 | 0.157 |
| **High** (N = 260) | | | | | | | | |
| Mode | 3 | 3 | 3 | 2 | 3 | 3 | 3 | 3 |
| Median | 3 | 3 | 3 | 2 | 3 | 3 | 3 | 3 |
| Mean | 3 | 2.846 | 2.331 | 2.154 | 2.677 | 2.869 | 2.635 | 2.819 |
| Std. Deviation | 0 | 0.439 | 0.855 | 0.669 | 0.523 | 0.349 | 0.603 | 0.415 |
| Coeff. of var. | 0 | 0.154 | 0.367 | 0.311 | 0.195 | 0.122 | 0.229 | 0.147 |

**Table 8**
Descriptives of the evaluation scores in three spectra (low = 1, neutral = 2, high = 3) in relation to the self-assessment reference score $A_S$ for the agreeableness dimension. Note that whenever more than one mode exists, only the first one is reported.

| Agreeableness | $A_S$ | $A_P$ | $H_A$ | $H_B$ | $GPT_{TL}$ | $GPT_{DTL}$ | $GPT_{LT}$ | $GPT_{DLT}$ |
|---|---|---|---|---|---|---|---|---|
| **Low** (N = 40) | | | | | | | | |
| Mode | 1 | 3 | 3 | 2 | 3 | 3 | 3 | 3 |
| Median | 1 | 3 | 3 | 2 | 3 | 3 | 3 | 3 |
| Mean | 1 | 2.2 | 2.325 | 2 | 2.725 | 2.725 | 2.65 | 2.75 |
| Std. Deviation | 0 | 0.992 | 0.859 | 0.784 | 0.554 | 0.599 | 0.736 | 0.63 |
| Coeff. of var. | 0 | 0.451 | 0.369 | 0.392 | 0.203 | 0.22 | 0.278 | 0.229 |
| **Neutral** (N = 196) | | | | | | | | |
| Mode | 2 | 3 | 3 | 2 | 3 | 3 | 3 | 3 |
| Median | 2 | 3 | 3 | 2 | 3 | 3 | 3 | 3 |
| Mean | 2 | 2.531 | 2.306 | 2.051 | 2.73 | 2.679 | 2.673 | 2.76 |
| Std. Deviation | 0 | 0.675 | 0.888 | 0.749 | 0.539 | 0.635 | 0.684 | 0.598 |
| Coeff. of var. | 0 | 0.267 | 0.385 | 0.365 | 0.197 | 0.237 | 0.256 | 0.217 |
| **High** (N = 384) | | | | | | | | |
| Mode | 3 | 3 | 3 | 2 | 3 | 3 | 3 | 3 |
| Median | 3 | 3 | 3 | 2 | 3 | 3 | 3 | 3 |
| Mean | 3 | 2.667 | 2.432 | 2.081 | 2.797 | 2.786 | 2.789 | 2.875 |
| Std. Deviation | 0 | 0.534 | 0.84 | 0.762 | 0.496 | 0.522 | 0.555 | 0.434 |
| Coeff. of var. | 0 | 0.2 | 0.345 | 0.366 | 0.177 | 0.187 | 0.199 | 0.151 |

**Table 9**
Descriptives of the evaluation scores in three spectra (low = 1, neutral = 2, high = 3) in relation to the self-assessment reference score $A_S$ for the conscientiousness dimension. Note that whenever more than one mode exists, only the first one is reported.

| Conscientiousness | $A_S$ | $A_P$ | $H_A$ | $H_B$ | $GPT_{TL}$ | $GPT_{DTL}$ | $GPT_{LT}$ | $GPT_{DLT}$ |
|---|---|---|---|---|---|---|---|---|
| **Low** (N = 52) | | | | | | | | |
| Mode | 1 | 2 | 2 | 2 | 2 | 2 | 3 | 3 |
| Median | 1 | 2 | 2 | 2 | 2 | 2 | 3 | 3 |
| Mean | 1 | 1.846 | 2.269 | 1.865 | 2.115 | 2.346 | 2.538 | 2.442 |
| Std. Deviation | 0 | 0.538 | 0.66 | 0.561 | 0.646 | 0.653 | 0.67 | 0.669 |
| Coeff. of var. | 0 | 0.291 | 0.291 | 0.301 | 0.306 | 0.278 | 0.264 | 0.274 |
| **Neutral** (N = 340) | | | | | | | | |
| Mode | 2 | 2 | 3 | 2 | 2 | 2 | 3 | 3 |
| Median | 2 | 2 | 2 | 2 | 2 | 2 | 3 | 3 |
| Mean | 2 | 2.459 | 2.45 | 2.059 | 2.138 | 2.432 | 2.659 | 2.665 |
| Std. Deviation | 0 | 0.522 | 0.585 | 0.476 | 0.505 | 0.568 | 0.581 | 0.579 |
| Coeff. of var. | 0 | 0.212 | 0.239 | 0.231 | 0.236 | 0.234 | 0.219 | 0.217 |
| **High** (N = 228) | | | | | | | | |
| Mode | 3 | 3 | 3 | 2 | 2 | 3 | 3 | 3 |
| Median | 3 | 3 | 3 | 2 | 2 | 3 | 3 | 3 |
| Mean | 3 | 2.632 | 2.57 | 2.158 | 2.162 | 2.544 | 2.772 | 2.789 |
| Std. Deviation | 0 | 0.483 | 0.522 | 0.498 | 0.425 | 0.533 | 0.461 | 0.45 |
| Coeff. of var. | 0 | 0.184 | 0.203 | 0.231 | 0.197 | 0.21 | 0.166 | 0.161 |

**Table 10**

Descriptives of the evaluation scores in three spectra (low = 1, neutral = 2, high = 3) in relation to the self-assessment reference score $A_S$ for the neuroticism dimension. Note that whenever more than one mode exists, only the first one is reported.

| Neuroticism | $A_S$ | $A_P$ | $H_A$ | $H_B$ | $GPT_{TL}$ | $GPT_{DTL}$ | $GPT_{LT}$ | $GPT_{DLT}$ |
|---|---|---|---|---|---|---|---|---|
| **Low** (N = 204) | | | | | | | | |
| Mode | 1 | 1 | 3 | 3 | 2 | 1 | 2 | 1 |
| Median | 1 | 2 | 3 | 2 | 2 | 1 | 2 | 2 |
| Mean | 1 | 1.765 | 2.422 | 2.186 | 1.828 | 1.48 | 1.985 | 1.667 |
| Std. Deviation | 0 | 0.732 | 0.768 | 0.759 | 0.655 | 0.616 | 0.791 | 0.692 |
| Coeff. of var. | 0 | 0.415 | 0.317 | 0.347 | 0.358 | 0.416 | 0.398 | 0.415 |
| **Neutral** (N = 216) | | | | | | | | |
| Mode | 2 | 3 | 3 | 2 | 2 | 1 | 2 | 1 |
| Median | 2 | 2.5 | 3 | 2 | 2 | 1 | 2 | 1 |
| Mean | 2 | 2.296 | 2.44 | 2.069 | 1.801 | 1.472 | 2.028 | 1.602 |
| Std. Deviation | 0 | 0.787 | 0.769 | 0.795 | 0.596 | 0.57 | 0.783 | 0.674 |
| Coeff. of var. | 0 | 0.343 | 0.315 | 0.384 | 0.331 | 0.387 | 0.386 | 0.421 |
| **High** (N = 200) | | | | | | | | |
| Mode | 3 | 3 | 3 | 2 | 2 | 1 | 2 | 1 |
| Median | 3 | 3 | 3 | 2 | 2 | 1 | 2 | 2 |
| Mean | 3 | 2.58 | 2.455 | 2.23 | 1.805 | 1.515 | 2.075 | 1.65 |
| Std. Deviation | 0 | 0.697 | 0.735 | 0.728 | 0.591 | 0.601 | 0.736 | 0.663 |
| Coeff. of var. | 0 | 0.27 | 0.3 | 0.326 | 0.327 | 0.397 | 0.355 | 0.402 |

**Table 11**

Descriptives of the evaluation scores in three spectra (low = 1, neutral = 2, high = 3) in relation to the self-assessment reference score $A_S$ for the openness dimension. Note that whenever more than one mode exists, only the first one is reported.

| Openness | $A_S$ | $A_P$ | $H_A$ | $H_B$ | $GPT_{TL}$ | $GPT_{DTL}$ | $GPT_{LT}$ | $GPT_{DLT}$ |
|---|---|---|---|---|---|---|---|---|
| **Low** (N = 8) | | | | | | | | |
| Mode | 1 | 1 | 3 | 2 | 2 | 2 | 2 | 3 |
| Median | 1 | 1.5 | 2.5 | 2 | 2 | 2 | 2 | 2.5 |
| Mean | 1 | 1.5 | 2.375 | 2.25 | 1.75 | 1.75 | 2.25 | 2.375 |
| Std. Deviation | 0 | 0.535 | 0.744 | 0.463 | 0.463 | 0.707 | 0.707 | 0.744 |
| Coeff. of var. | 0 | 0.356 | 0.313 | 0.206 | 0.265 | 0.404 | 0.314 | 0.313 |
| **Neutral** (N = 296) | | | | | | | | |
| Mode | 2 | 2 | 2 | 2 | 2 | 2 | 3 | 2 |
| Median | 2 | 2 | 2 | 2 | 2 | 2 | 3 | 2 |
| Mean | 2 | 2.243 | 2.199 | 2.01 | 2.027 | 1.794 | 2.541 | 2.213 |
| Std. Deviation | 0 | 0.541 | 0.667 | 0.39 | 0.216 | 0.554 | 0.587 | 0.678 |
| Coeff. of var. | 0 | 0.241 | 0.303 | 0.194 | 0.107 | 0.309 | 0.231 | 0.306 |
| **High** (N = 316) | | | | | | | | |
| Mode | 3 | 3 | 2 | 2 | 2 | 2 | 3 | 2 |
| Median | 3 | 3 | 2 | 2 | 2 | 2 | 3 | 2 |
| Mean | 3 | 2.684 | 2.247 | 2.044 | 2.041 | 1.886 | 2.582 | 2.345 |
| Std. Deviation | 0 | 0.492 | 0.673 | 0.38 | 0.229 | 0.574 | 0.593 | 0.651 |
| Coeff. of var. | 0 | 0.183 | 0.3 | 0.186 | 0.112 | 0.305 | 0.23 | 0.278 |